\documentclass[twocolumn,showpacs,pra]{revtex4}
\usepackage{graphicx}
\newcommand{\ket}[1]{|{#1}\rangle}			
\newcommand{\bra}[1]{\langle{#1}|}			
\newcommand{\ketbra}[3]{\ket{#1}_{#3}\bra{#2}}

\begin{document}
\title{Unconditional Security of the Bennett 1992 quantum key-distribution over lossy and noisy channel}

\author{Kiyoshi Tamaki$^{1}$}
\email{tamaki_kiyoshi@soken.ac.jp}
\author{Norbert L\"{u}tkenhaus$^{2}$}
%\homepage{http://www.soken.ac.jp/quantum/index.html}
\affiliation{$^{1}$The Graduate University for  Advanced Studies
(SOKENDAI), Hayama, Kanagawa, 240-0193, Japan,\\ $^{2}$Institut f\"{u}r Theoretische Physik I,
Universit\"{a}t Erlangen-N\"{u}rnberg, Staudtstr. 7/B3, 91058 Erlangen, Germany
}


%\date{\today}

\begin{abstract} 
We show that the security proof of the Bennett 1992 protocol over loss-free channel in (K.~Tamaki, M.~Koashi, and N.~Imoto, Phys.~Rev.~Lett. {\bf{90}}, 167904 (2003)) can be adapted to accommodate loss. We assumed that Bob's detectors discriminate between single photon states on one hand and vacuum state or multi-photon states on the other hand.

\end{abstract}

\maketitle
%%%%%%%%%%%%%Text%%%%%%%%%%

\section{Introduction}
Quantum key distribution (QKD) is one of the most interesting topics in quantum information processing, which allows the sender (Alice) and the receiver (Bob) to share a secret key with negligibly leakage of information to an eavesdropper (Eve). This task is accomplished by making use of the properties of quantum mechanics, such as no-cloning theorem \cite{noclone}, indistinguishability of nonorthogonal quantum states, entanglement, and so on \cite{peres}. The B92 protocol \cite{B92}, where only two nonorthogonal states are used, is thought to be secure due to the fact that Eve cannot clone and distinguish the states deterministically.

Although the qualitative reason of the security is simple, it is quite hard to show the quantitative security against so-called coherent attack, which is the most general attack allowed by quantum mechanics. In \cite{lc98,sp00}, the security of the BB84 protocol \cite{BB84} is proven by showing the relationship between QKD and other important protocols in quantum information, such as the entanglement distillation protocol (EDP) \cite{EPP} and the Calderbank-Shor-Steane (CSS) quantum error correcting codes \cite{CSS}. Recently in \cite{tki03}, the unconditional security of the B92 protocol has been proved by relating the B92 protocol to an entanglement distillation protocol (EDP) \cite{EPP} initiated by local filtering \cite{Gisin96}. The assumptions of the proof are that Alice and Bob's devises are perfect and the quantum channel is loss-free. To study the unconditional security of B92 protocol over a lossy and noisy channel is interesting from the point of view of the quantum information theory, since the channel loss directly affects the security of the B92 protocol. The proof based on the equivalence between the security and EDPs gives us an idea how to distill the EPR state from states that arise from the lossy and noisy channel. 

In this paper, we study the unconditional security of the B92 protocol. We assume that Alice has a perfect single photon source, Bob's polarization measurement is perfect and he has a perfect single photon counter that discriminates between single photon states on one hand and vacuum state or multi-photon states on the other hand. Under these assumptions, we show the unconditional security of the B92 protocol. The proof follows the idea in \cite{tki03}. This paper is organized as follows. In Sec.~\ref{review}, we propose the protocol based on EDP with local filtering and CSS codes \cite{sp00}, and we reduce this to the B92 protocol. In Sec.~\ref{estimation} we give the formula for the error estimations for EDP based on CSS code. Finally, in Sec.~\ref{conclusion} we give the examples of the resulting performance, and the summary and discussion follow.

%%%%%%%%%%%%%%%%%%%%%%%%%%%%%%%%
%%%%%%%%%%%%%%%%%%%%%%%%%%%%%%%%
\section{Protocol based on EDP and its reduction to the B92 protocol}\label{review} 
%%%%%%%%%%%%%%%%%%%%%%%%%%%%%%%%
%%%%%%%%%%%%%%%%%%%%%%%%%%%%%%%%

In this section, we introduce the protocol that will turn out to be secure and will reduce to the B92 protocol. In this protocol, Alice first prepares a two-qubit nonmaximally entangled states that can be written as 
\begin{eqnarray}
\ket{\Psi}_{\rm{AB}}=\frac{1}{\sqrt{2}}\left(\ket{0_{z}}_{\rm{A}}\ket{\varphi_0}_{\rm{B}}+\ket{1_{z}}_{\rm{A}}\ket{\varphi_1}_{\rm{B}}\right)
\end{eqnarray}
and sends system B to Bob. Here $\ket{\varphi_j}\equiv\beta\ket{0_x}+(-1)^j\alpha\ket{1_x}$, ($j=0,1$), where $0<\alpha< 1/\sqrt{2}$, $\beta\equiv\sqrt{1-\alpha^2}$, and $\{\ket{0_x},\ket{1_x}\}$ is a basis ($X$-basis) of the qubit. $X$-basis and $Z$-basis is related through the relationship as $\ket{j_z}=[\ket{0_x}+(-1)^j\ket{1_x}]/\sqrt{2}$. 

After receiving the state, Bob first performs the QND measurements that can be described by the POVM (positive operator valued measure) \cite{peres}
\begin{eqnarray}
Q_{\rm{s}}&=&\Pi_{\rm{s}} \\
Q_{\rm{v}}&=&\openone-\Pi_{\rm{s}}
\end{eqnarray}
where $\Pi_{\rm{s}}$ is the projector onto the single photon space and ``v'' means the detection of vacuum state or multi-photon states. Thanks to this QND measurements, even if Eve sends multi-photon states or vacuum state to Bob, Alice and Bob can work only on the qubit pairs, provided that the outcome is ``s''. We also assume that, after this measurement, if the outcome is ``s'', Bob performs the ``local filtering operation'' \cite{Gisin96} on qubit B, described by the Kraus operator
\begin{eqnarray}
A_{\rm{fil}}\equiv\frac{\gamma}{\beta}\left(\alpha\ket{0_x}_{\rm B}\bra{0_x}+\beta\ket{1_x}_{\rm B}\bra{1_x}\right)\,,
\label{coefficient}
\end{eqnarray}
where $\gamma\in (0,1]$. Note that if the channel is loss-free and noiseless or Eve does nothing, this operation yields the maximally entangled state (EPR state) $(\ket{0_{x}}_{\rm{A}}\ket{0_{x}}_{\rm{B}}+\ket{1_{x}}_{\rm{A}}\ket{1_{x}}_{\rm{B}})/\sqrt{2}$ with probability $2\alpha^2\gamma^2$ due to the fact that the initial state is also written as $\ket{\Psi}_{\rm{AB}}=\beta\ket{0_{x}}_{\rm{A}}\ket{0_{x}}_{\rm{B}}+\alpha\ket{1_{x}}_{\rm{A}}\ket{1_{x}}_{\rm{B}}$. Thus, the filtering operation with $\gamma=1$ is optimum one in this case. When channel is lossy and noisy, or Eve is present, the filtered states may include bit errors, represented by the subspace spanned by $\{\ket{0_{z}}_{\rm{A}}\ket{1_{z}}_{\rm{B}},\ket{1_{z}}_{\rm{A}}\ket{0_{z}}_{\rm{B}}\}$, and a phase error, represented by the subspace spanned by $\{\ket{0_{x}}_{\rm{A}}\ket{1_{x}}_{\rm{B}},\ket{1_{x}}_{\rm{A}}\ket{0_{x}}_{\rm{B}}\}$. In order to distill the EPR pairs, Alice and Bob run the EDP based on CSS code \cite{sp00} and after this EDP Alice and Bob measure each in the $Z$-basis of the EPR pairs so that they obtain a secure key. Note that the QND measurement followed by the filtering $A_{\rm{fil}}$ and $Z$-basis measurement is equivalent to a measurement ${\cal M}_{\rm B92}^{L,\gamma}$ described by $F_{0}^{\gamma}=\gamma^2\ket{\overline{\varphi_1}}\bra{\overline{\varphi_1}}/(2 \beta^2)$, $F_{1}^{\gamma}=\gamma^2\ket{\overline{\varphi_0}}\bra{\overline{\varphi_0}}/(2 \beta^2)$, $F_{\rm null}^{\gamma}=\Pi_{\rm{s}}-F_{0}^{\gamma}-F_{1}^{\gamma}$, and $F_{\rm{v}}=\openone -F_{0}^{\gamma}-F_{1}^{\gamma}-F_{\rm null}^{\gamma}$, where $\ket{\overline{\varphi_j}}\equiv\alpha\ket{0_x}-(-1)^j\beta\ket{1_x}$. Thus ${\cal M}_{\rm B92}^{L,\beta}$ is the same measurement as in the B92 with perfect photon detector, and ${\cal M}_{\rm B92}^{L,\gamma}$ corresponds to the generalized B92 protocol.  For convenience, let ${\cal M}_{\rm B92}^{\gamma}$ be the measurements described by the set of the POVM elements $\{F_{0}^{\gamma},F_{1}^{\gamma},F_{\rm null}^{\gamma}\}$. The following is the protocol that we want to be shown secure and reduce to the generalized B92 protocol.

{\em Protocol:} 
(1) Alice prepares the state $\ket{\Psi}_{\rm{AB}}$, and she sends the system B to Bob over a quantum channel.  
(2) Bob performs the QND measurement described by $\{Q_{\rm{s}}, Q_{\rm{v}}\}$, and he publicly announces the outcome. They take notes of the outcome.
(3) Alice and Bob repeats the step (1) and (2) $2N$ times. Let $2LN$ be the number of outcomes that resulted in ``v''.
(4) Alice and Bob discard all the pairs whose outcomes of the QND measurement are ``v''. By public discussion, Alice and Bob randomly permute the positions of the remaining $2(1-L)N$ pairs. 
(5) For the first $(1-L)N$ pairs (check pairs), Alice measures system A in the $Z$-basis, and Bob performs measurement ${\cal M}_{\rm B92}^{\gamma}$ on his system. By public discussion, they determine the number $n_{\rm err}$ of errors in which Alice found $0$ and Bob's outcome was $1$, or Alice found $1$ with Bob's outcome $0$.
(6) For the second $(1-L)N$ pairs (data pairs), Bob performs the filtering $A_{\rm{fil}}$ on each of his qubits, and announces the total number $n_{\rm fil}$ and the positions of the qubits that have passed the filtering.
(7) From $L$, $n_{\rm err}$ and $n_{\rm fil}$, they estimate an upper
bound for the number of bit errors $n_{\rm bit}$, and an upper bound for the number of phase errors $n_{\rm ph}$, in the $n_{\rm fil}$ pairs. If these bounds are too large, they abort the protocol.
(8) They run an EDP that can produce $n_{\rm key}$ nearly perfect EPR pairs if the estimation is correct.
(9) Alice and Bob each measures the EPR pairs in $Z$-basis to obtain a shared secret key.

In the following, we will reduce our protocol to the B92 protocol. Before this reduction, we should briefly mention what Alice and Bob do experimentally in the B92 protocol. In the B92 protocol, Alice randomly prepares the single photon polarization in the state $\ket{\varphi_0}$ or $\ket{\varphi_1}$ and sends the state to Bob. After receiving the state, Bob performs measurement described by ${\cal M}_{\rm B92}^{L,\gamma}$. When $\gamma=\beta$ and Bob detects a single photon state, this measurement effectively performs a polarization measurement on the single photon state in a basis randomly chosen from $\{\ket{\varphi_0}, \ket{\overline\varphi_0}\}$ and $\{\ket{\varphi_1}, \ket{\overline\varphi_1}\}$. In the case where $\gamma=1$ and Bob detects a single photon state, this measurement effectively performs a general measurement which describes the optimum unambiguous state discrimination between $\ket{\varphi_0}$ and $\ket{\varphi_1}$ \cite{unb}.

Thanks to Bob's perfect detector, if the outcome of the QND measurement is ``s'', then Bob's states are projected onto the single photon space. This projection requires two more steps in our protocol compared to the {\em Protocol 1} in \cite{tki03}. The benefit of this projection is that for the states projected onto qubit-pairs Alice and Bob can use the one-way EDP based on CSS code \cite{sp00}. 
In the context of QKD, this EDP allows Alice to perform $Z$-basis measurements immediately after she has prepared the state $\ket{\Psi}_{\rm{AB}}$ and Bob to perform $Z$-basis measurements immediately after he has performed the filtering. Now Bob's measurement is described by ${\cal M}_{\rm B92}^{\gamma}$. Since Bob performs the QND measurement before the measurement ${\cal M}_{\rm B92}^{\gamma}$, his measurement is equivalent to the measurement ${\cal M}_{\rm B92}^{L, \gamma}$. As a whole, the protocol has been reduced into the protocol where  Alice prepares and sends $\ket{\varphi_0}$ or $\ket{\varphi_1}$ randomly and Bob performs the ${\cal M}_{\rm B92}^{L,\gamma}$ measurement. This completes the reduction of our protocol into the generalized B92 protocol over lossy and noisy quantum channel.

Due to this equivalence, it is sufficient to prove the security of the protocol for the security analysis of the B92. The most important part in the quantitative security analysis of our protocol is to estimate phase errors and bit errors in the step (7). If this estimation is not exponentially reliable, then in the protocol Alice and Bob cannot distill the state sufficiently close to EPR state so that the protocol is not secure \cite{lc98, sp00}. In the next section, we will give the formula to estimate the bit errors and phase errors.

%%%%%%%%%%%%%%%%%%%%%%%%%%%%%%%%%%%%%%%%%%%
%%%%%%%%%%%%%%%%%%%%%%%%%%%%%%%%%%%%%%%%%%%
\section{Phase error estimation from the bit error}\label{estimation} 
%%%%%%%%%%%%%%%%%%%%%%%%%%%%%%%%%%%%%%%%%%%
%%%%%%%%%%%%%%%%%%%%%%%%%%%%%%%%%%%%%%%%%%%

In this section, we consider the error estimations in the step (7) of the protocol. Thanks to Bob's perfect detector, the filtered states are qubit states, which allows us to directly apply some inequalities from \cite{tki03}. 

First, note that bit errors $n_{\rm bit}$ in the data pairs could be determined if Alice measures those pairs in $Z$-basis, Bob performs measurement ${\cal M}_{\rm B92}^{\gamma}$ onto those pairs, and they compare their results. Thus, the bit error rate $n_{\rm err}$ in the check pairs and bit errors $n_{\rm bit}$ in the data pairs are determined by the same measurements. Furthermore thanks to the random permutation in step (4), we can apply the classical probability estimation so that we obtain
\begin{equation}
|n_{\rm bit}-n_{\rm err}|\le (1-L)N\epsilon_1.
\label{ineq1}
\end{equation}
For any strategy by Eve, the probability of violating this inequality is asymptotically less than $\exp\left[-(1-L)N \epsilon_1^{2}\right]$. 

Since the POVM elements that give the bit error rate on the filtered states $\Pi_{\rm bit}$ and phase error rate on the filtered states $\Pi_{\rm ph}$ do not commute \cite{POVMelement}, the classical probability argument cannot be applied to the estimation of the phase errors from the bit errors. For the estimation, let us perform the gedanken measurements in the $\ket{i_x}_{\rm{A}}\ket{j'_x}_{\rm{B}}$ basis $(i=0,1,\,\,j'=0,1,{\rm{v}})$, and let the numbers $n_{ij'}$ of the pairs found in the state $\ket{i_x}_{\rm{A}}\ket{j'_x}_{\rm{B}}$. Here we define $n_{i\rm{v}}$ as the outcome where Alice obtains the bit value $i$ and Bob obtains vacuum or multi-photon states. As far as we consider the measurement outcomes of the POVM elements that are diagonal with respect to $\ket{i_x}_{\rm{A}}\ket{j'_x}_{\rm{B}}$ basis, we are allowed to apply the classical argument for the estimation. Thus, we have the following inequalities,
\begin{eqnarray}
\left|\alpha^2(n_{00}+n_{10})+\beta^2(n_{01}+n_{11})-\frac{\beta^2}{\gamma^2}n_{\rm fil}\right|\le (1-L)N\epsilon_2 
\label{ineq2} \\
\left|\alpha^2n_{10}+\beta^2n_{01}-\frac{\beta^2}{\gamma^2}n_{\rm ph}\right|\le (1-L)N\epsilon_3,
\label{ineq3}
\end{eqnarray}
which are violated with probability asymptotically less than $\exp\left[-2(1-L)N\epsilon_{2}^2\right]$ and $\exp\left[-2(1-L)N\epsilon_{3}^2\right]$, respectively. Similarly, since neither the noisy channel nor Eve can touch the qubits held by Alice, we have
\begin{equation}
|\alpha^2N-(n_{10}+n_{11}+n_{1{\rm{v}}})|
 \le N\epsilon_4\, ,
\label{ineq4}
\end{equation}
with probability of violation asymptotically less than $\exp(-2N\epsilon_{4}^2)$.

In order to represent the bit error in the check pairs in terms of gedanken measurements, we introduce the global gedanken measurements in the basis  $\ket{\Gamma_{ij}}$ $(i=0,1,\,\,j=0,1)$ and the number of the corresponding measurement outcomes $m_{ij}$, where $\ket{\Gamma_{ij}}\equiv(-1)^{ij}\beta\ket{i_x}_{\rm{A}}\ket{j_x}_{\rm{B}}+(-1)^{i(j+1)}\alpha\ket{(i+1)_x}_{\rm{A}}\ket{(j+1)_x}_{\rm{B}}$. Here the summation is taken modulo 2. Since the POVM element for the bit error is diagonal with respect to the $\ket{\Gamma_{ij}}$ basis \cite{POVMelement},
we have
\begin{eqnarray}
\left|(m_{11}+m_{01})/2-\frac{\beta^2}{\gamma^2}n_{\rm err}\right|
 \le (1-L)N\epsilon_5,
\label{ineq5}
\end{eqnarray}
which is violated with probability asymptotically less than $\exp(-2(1-L)N\epsilon_{5}^2)$. 

Since $\{\ket{\Gamma_{01}},\ket{\Gamma_{10}}\}$ and $\{\ket{0_x}_{\rm{A}}\ket{1_x}_{\rm{B}}, \ket{1_x}_{\rm{A}}\ket{0_x}_{\rm{B}} \}$ span the same subspace, we can relate $m_{10}+m_{01}$ and $n_{10}+n_{01}$ by the classical probability estimate as
\begin{equation}
|(m_{10}+m_{01})-(n_{10}+n_{01})|
 \le (1-L)N\epsilon_6,
\label{ineq6}
\end{equation}
which is violated with probability asymptotically less than 
$\exp\left[-L(1-L)N \epsilon_6^{2}\right]$. For the estimate of the phase error rate in the data qubit from the bit error rate in the check qubit, we need to relate $m_{01}/(m_{01}+m_{10})$ to $n_{01}/(n_{01}+n_{10})$. This relation cannot be obtained by classical arguments, since $\ket{\Gamma_{01}}$ and $\ket{0_x}_{\rm{A}}\ket{1_x}_{\rm{B}}$ are nonorthogonal. In \cite{tki03}, the authors consider the qubits that are invariant under the random permutation in the limit of large number. Then they showed that when we consider the measurement outcomes onto such qubits, we are allowed to regard the qubits as if they arose from the independently and identically distributed quantum source. Their argument can be directly applied to our case, since in step (4) of the protocol we have also applied the random permutation. Thus, we can make use of 
the exponentially reliable inequality derived in \cite{tki03} as
\begin{equation} 
\sin^2(\theta_l-\theta)-\epsilon_7\le \sin^2\phi_l \le
\sin^2(\theta_l+\theta)+\epsilon_8
\label{lastineq}
\end{equation} 
for $l=0,1$, where all the angles are defined in $[0,\pi/2]$ 
by the relations
$n_{11}/(n_{11}+n_{00})=\sin^2\theta_0$,
$n_{01}/(n_{01}+n_{10})=\sin^2\theta_1$, 
$m_{11}/(m_{11}+m_{00})=\sin^2\phi_0$,
$m_{01}/(m_{01}+m_{10})=\sin^2\phi_1$, 
and
$\alpha^2=\sin^2\theta$.

The Eqs.~(\ref{ineq1})--(\ref{lastineq}), together with the relation $\sum n_{ij'}=N$ and $\sum m_{ij}=(1-L)N$ can be used to derive the formula for the estimation of the phase error from bit error. In the following we set $\epsilon_i\rightarrow 0$, $(i=1, 2, \ldots 8)$ by taking the limit $N\rightarrow\infty$ so that Eqs.~(\ref{ineq1})--(\ref{ineq6}) are now linear equations.

We first consider the special case for the B92 where bit error rate in the filtered qubits is zero while the phase error rate in the qubits is not zero. Eve can accomplish this by using channel loss. From $n_{\rm bit}=0$, i.e., $m_{11}=m_{01}=0$, one can easily find that for $n_{\rm fil}/N=2\alpha^2\gamma^2(1-L)$
\begin{equation}
{\rm Max} \left(\frac{n_{\rm ph}}{n_{\rm fil}}\right) \left\{
\begin{array}{ll}
=\frac{\alpha^2}{\beta^2-\alpha^2}\frac{L}{1-L} & \mbox{for $L\le\beta^2-\alpha^2$} \\ \\ \ge1/2 & \mbox{for $L>\beta^2-\alpha^2$}
\end{array}
\right. \; .
\end{equation}
The limiting loss rate $\beta^2-\alpha^2$ coincides with the case where Eve employs optimum unambiguous state discrimination measurement \cite{unb}. In this attack, if she obtains the conclusive outcome, then she sends the corresponding state to Bob, otherwise she does not. 

Next we consider the general case. The relationship can be obtained by solving the equations that we have mentioned just above. After straight forward calculations, we find the implicit constraint
\begin{eqnarray}
\frac{\beta^2}{\gamma^2}\left|n_{\rm fil}-2n_{\rm err}\right|\nonumber\le N\alpha\,\beta f(x),
\label{ineqlimit}
\end{eqnarray}
where 
\begin{eqnarray}
f(x)&\equiv&\sqrt{\left(x-L_1\right)^2-\left(-\Delta+\frac{L_1}{\alpha^2-\beta^2}\right)^2}\nonumber \\
&+&\sqrt{\left(1-x-L_0\right)^2-\left(\beta^2-\alpha^2-\Delta+\frac{L_0}{\alpha^2-\beta^2}\right)^2}\nonumber\\
\Delta&\equiv& \left[\left(\beta^2/\gamma^2\right)n_{\rm fil}/N-2\alpha^2\beta^2\right]/(\beta^2-\alpha^2) \nonumber\\
x&\equiv&\left(\beta^2/\gamma^2\right) 2 n_{\rm ph}/N-(\beta^2-\alpha^2)\Delta \nonumber\\
L_0&\equiv&\alpha^2 n_{1{\rm v}}/N+\beta^2 n_{0{\rm v}}/N \nonumber\\
L_1&\equiv&\alpha^2 n_{0{\rm v}}/N+\beta^2 n_{1{\rm v}}/N
\end{eqnarray} 
Here, $x-L_1$ and $1-x-L_0$ are equal to $(n_{10}+n_{01})/N$ and $(n_{00}+n_{11})/N$ respectively so that they must be positive. Combining these positivity constraints with the positivities inside the square roots, we have $L_1+|-\Delta+\frac{L_1}{\alpha^2-\beta^2}|\le x \le 1-L_0-|-\Delta-\alpha^2+\beta^2+\frac{L_0}{\alpha^2-\beta^2}|$. We can easily confirm that these equations are the same as equations in \cite{tki03} where $\gamma=\beta$ and the channel has no loss i.e., $L_0=L_1=0$. Eq.~(\ref{ineqlimit}) gives us the possible parameter set $\{n_{\rm err}, n_{\rm ph}\}$
 for fixed $\alpha$, $\gamma$, $L_{0}$, $L_{1}$, $L$, and $n_{\rm fil}$. By taking the optimum ratio between $L_{0}$ and $L_{1}$ with respect to $n_{\rm ph}$ for given $n_{\rm err}$, $n_{\rm fil}$ and $L$, we obtain an upper bound $\overline{n}_{\rm ph}$ on the number of phase errors $n_{\rm ph}$, as a function of the observed values $n_{\rm err}$, $n_{\rm fil}$ and $L$, and the chosen parameters $\alpha$ and $\gamma$.

%%%%%%%%%%%%%%%%%%%%%%%%%%%%%%%%%%%%%%%%%%%
%%%%%%%%%%%%%%%%%%%%%%%%%%%%%%%%%%%%%%%%%%%
\section{Examples and Conclusion}\label{conclusion}
%%%%%%%%%%%%%%%%%%%%%%%%%%%%%%%%%%%%%%%%%%%
%%%%%%%%%%%%%%%%%%%%%%%%%%%%%%%%%%%%%%%%%%%

In this section, at first we give the examples of the phase error estimation and resulting performance of the B92 protocol. 

\begin{figure}[tbp]
\begin{center}
 \includegraphics[scale=0.5]{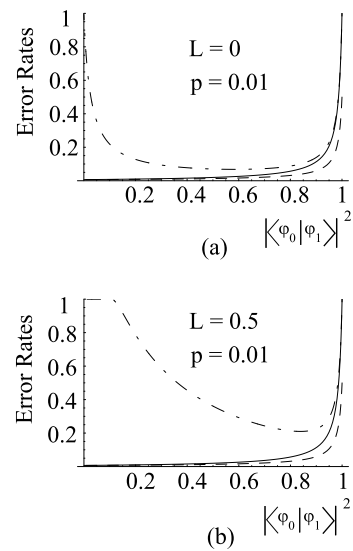}
\end{center}
 \caption{\setlength{\baselineskip}{0mm} The estimated phase error rate $\overline{n}_{\rm ph}/n_{\rm fil}$ (dot-dashed line), the actual phase error rate (solid line), and bit error rate (dotted line) in case of $p=0.01$. Diagram (a) is for $L=0$ and (b) is for $L=0.5$. We assume that quantum channel is a depolarizing channel with loss.
\label{phase}}
\end{figure}

To illustrate the accuracy of the phase error estimation and security performance of the B92 protocol, we assign values to the observable data as they would arise from a depolarizing quantum channel with loss, i.e., the state $\rho$ evolves into $L\ket{V}\bra{V}+(1-L)\left[(1-p)\rho+\sum_{i=x,y,z}\sigma_{i}\rho\sigma_{i}\right]$, where $\ket{V}$ is the vacuum state, $L$ is the loss rate of quantum channel, $p$ is the depolarizing rate, and $\sigma_{i}$ is the Pauli matrix. In Fig.~\ref{phase}, we show the upper bound of the estimated phase error rate, the actual phase error rate, and bit error rate for the case (a) $p=0.01, \,\, L=0$, and (b) $p=0.01, \,\, L=0.5$. These figures are independent of $\gamma$, since we plot the phase error rates normalized by $n_{\rm fil}$ so that the pre-factor $\beta^2/\gamma^2$ is cancelled. In each figure, it is seen that when $|\langle{\varphi_0}|{\varphi_1}\rangle|^2$ becomes smaller, the estimation of the phase errors becomes poorer. On the other hand, larger values of $|\langle{\varphi_0}|{\varphi_1}\rangle|^2$ make the signal more vulnerable to noise and loss. Furthermore, due to the loss, the estimation in (b) is poorer than that in (a), especially in the region for small $|\langle{\varphi_0}|{\varphi_1}\rangle|^2$.

%%%%%%%%%%%%%%%%%%%%%%%%%%%%%%%figure1
%%%%%%%%%%%%%%%%%%%%%%%%%%%%%%%
%%%%%%%%%%%%%%%%%%%%%%%%%%%%%%
%%%%%%%%%%%%%%%%%%%%%%%%%%%%%%

\begin{figure}[tbp]
\begin{center}
 \includegraphics[scale=0.5]{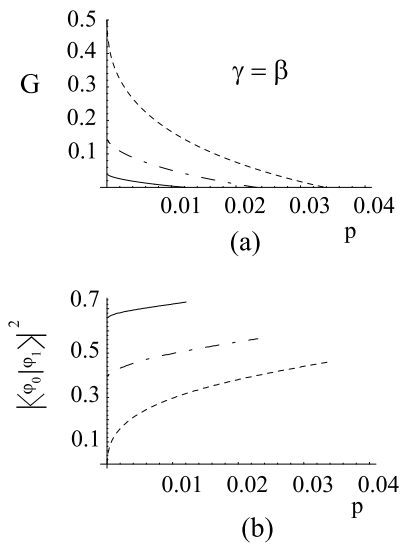}
\end{center}
 \caption{\setlength{\baselineskip}{0mm} (a) The optimum key generation rate $G$ for the case $\gamma=\beta$ and (b) the optimum value of $|\langle{\varphi_0}|{\varphi_1}\rangle|^2$. Here we assume that the quantum channel is the depolarizing channel with loss. The dotted line, the dot-dashed, and solid line represent the case where $L=0$, $L=0.2$, and $L=0.5$ respectively.
\label{keygain}}
\end{figure}

Since Alice and Bob apply random permutation in the protocol, the achievable length of the final key is given by $n_{\rm key}=(\gamma^2/\beta^2) n_{\rm{fil}}[1-h(n_{\rm bit}/n_{\rm fil})-h(\overline{n}_{\rm ph}/ n_{\rm fil})]$  \cite{tki03,mh03} , when $\overline{n}_{\rm ph}/n_{\rm fil}\le 1/2$. Here $h(p)\equiv-p\log_{2}p-(1-p)\log_{2}(1-p)$. The positivity of $n_{\rm key}$ means that Alice and Bob generate a secure key. 
In Fig.~\ref{keygain} (a), we have numerically calculated the key generation rate $G\equiv n_{\rm key}/N$ in the case of $\gamma=\beta$ for the fixed value of $|\langle{\varphi_0}|{\varphi_1}\rangle|^2$, $L$ and $p$, and plot $G$ optimized over $|\langle{\varphi_0}|{\varphi_1}\rangle|^2$ as a function of $p$. The optimum value of $|\langle{\varphi_0}|{\varphi_1}\rangle|^2$ is shown in Fig.~\ref{keygain} (b). Since the optimization of the ratio between $L_{0}$ and $L_{1}$ is complicated, we have optimized the ratio numerically. From Fig.~\ref{keygain} (a), it is seen that the B92 protocol as described here is secure up to $p\sim 0.034$ (in the case of $L=0$), $p\sim 0.023$ (in the case of $L=0.2$), and $p\sim 0.012$ (in the case of $L=0.5$). The use of the general B92 measurement, i.e., $\gamma=1$ yields the higher secret key gain, however this does not change the cutoff of the gain nor the optimum angle in Fig.~\ref{keygain} (b). This is again because of the fact that the cutoff is determined by the error rate normalized by $n_{\rm fil}$. 

In the BB84 protocol or 6-state protocol \cite{b98}, thanks to the randomly chosen conjugate bases, phase error estimation is more precise than that in the B92 protocol. The poor estimation in the B92 makes this protocol more vulnerable to eavesdropping. Actually, it is known that the BB84 protocol and 6-state protocol are secure up to $p\sim 0.165$ \cite{sp00} and $p\sim 0.1905$ \cite{l01} corresponding to bit error rates of $11\%$ and $12.7 \%$, respectively.

%%%%%%%%%%%%%%%%%%%%%%%%%%%%%%%figure2
%%%%%%%%%%%%%%%%%%%%%%%%%%%%%%%
%%%%%%%%%%%%%%%%%%%%%%%%%%%%%%
%%%%%%%%%%%%%%%%%%%%%%%%%%%%%%

In summary, we have shown that the security proof of the B92 over loss-free channel \cite{tki03} can be adapted to lossy and noisy channels assuming Bob's detector that discriminates between single photon states on one hand and vacuum state or multi-photon states on the other hand. We have derived the general formula for the estimation of the phase errors from bit errors. We have also shown that the boundary loss rate at which Eve can induce phase errors freely without inducing bit errors coincides with the case where Eve employs the optimum unambiguous state discrimination measurement. Finally, we have shown the examples of the security assuming the quantum channel is the depolarizing quantum channel with loss. 

It is an interesting problem to compare our results to the case where Eve performs only individual attack. For the B92 with loss, the security is estimated in \cite{tki03-2}. However in that paper, it is assumed that Alice and Bob employ error discarding protocol so that we cannot compare our result to their result directly. We leave this problem for future studies. Another interesting problem is to consider the case where Bob's detector is not perfect. In this case, some filtered states may remain vacuum state or multi-photon states so that our proof cannot be directly applied. The security analysis of the B92 with coherent state is another interesting problem. We hope our study is helpful for solving these problems and we leave them as future studies.

\section{Acknowledgement}
K.T appreciates the warmth and hospitality of the Quantum Information Theory Group, Universit\"{a}t Erlangen-N\"{u}rnberg in Germany during his stay. We thank to Marcos Curty, Peter van Loock, Philippe Raynal, and Aska Dolinska for helpful discussions. This work has been supported in part by the Academic Frontiers Student Exchange Promotion Program under Grant No.~60 by the Ministry of Education, Culture, Sports, Science and Technology, Japan, and by the DFG under the Emmy-Noether programme.

%\end{multicols}{2}

\begin{thebibliography}{99}
\bibitem{POVMelement} $\Pi_{\rm{bit}}$ is given by $(\ketbra{\Gamma_{11}}{\Gamma_{11}}{}+\ketbra{\Gamma_{01}}{\Gamma_{01}}{})/2$ (see the text for the definition of $\ket{\Gamma_{ij}}$), while $\Pi_{\rm{Ph}}$ is given by $\alpha^2\ket{1_x}_{\rm{A}}\bra{1_x}\otimes\ket{0_x}_{\rm{B}}\bra{0_x}+\beta^2\ket{0_x}_{\rm{A}}\bra{0_x}\otimes\ket{1_x}_{\rm{B}}\bra{1_x}$.
\end{thebibliography}
\end{document}